\documentstyle[12pt]{article}
\input{epsf}
\begin{document}
\hfill April 1999
\begin{center}

\vspace{32pt}

  { \large \bf Some Speculations on the Gauge Coupling \\in the 
AdS/CFT Approach }

\end{center}

\vspace{18pt}

\begin{center}
{P. Olesen}\\
 { The Niels Bohr Institute, Blegdamsvej 17,\\ 
~~DK-2100 Copenhagen \O, Denmark. \\e-mail: {\tt polesen@nbi.dk}}
\end{center}

\vspace{18pt}

\begin{center}

{\bf Abstract}

\end{center}

\bigskip
We propose the principle that the scale of the glueball masses in the AdS/CFT 
approach to QCD should be set by the square root of the string 
tension.  It then turns out that the strong bare coupling runs
logarithmically with the ultraviolet cutoff $T$ if first order
world sheet fluctuations are included. We also point out that
in the end, when all corrections are included, one should obtain an
equation for the coupling running with $T$ which has some similarity with
the equation for the strong bare coupling.

\vfill

\newpage

The remarkable duality between supersymmetric $SU(N)$ gauge theory and 
type IIB string theory
in anti de-Sitter space times a compact space \cite{Mal} and its 
finite temperature generalization to non-supersymmetric gauge
theories \cite{Witten} have been much discussed. It was pointed out
by Gross and Ooguri \cite{gr} that the four dimensional non-supersymmetric 
theory constructed this way corresponds to four dimensional large $N$
QCD only in a limit where the temperature $T$ approaches infinity
and at the same time the coupling $\lambda=g_{YM}^2N$ goes to zero. More
precisely, the limits to be taken are \cite{gr}
\begin{equation}
T\rightarrow\infty~~~{\rm and}~~~\lambda\rightarrow\frac{b}
{\ln(T/\Lambda_{QCD})}.
\label{scaling}
\end{equation}
Here $\Lambda_{QCD}$ is a renormalization group invariant, so
it should not change by simultaneous changes of $T$ and $\lambda$.
To actually take these limits is not feasible at the moment, since the 
supergravity approximation breaks down for $\lambda\rightarrow 0$. 

In the supergravity approximation to the AdS/CFT approach the temperature 
$T$ plays the role of an
ultraviolet cutoff, and the coupling $\lambda_T=g_{YM}^2N\rightarrow\infty$
is the bare coupling at the scale $T$. The
string tension in the saddle point approximation is then proportional 
to $\lambda_T T^2$ \cite{Witten,gr,tension}. Here
the coupling is, however, an arbitrary parameter (as long as it is large),
and does not seem run with the scale $T$. 

This causes a problem \cite{jeff} concerning the comparison of the string 
tension and the glueball masses in the strong coupling limit 
\cite{Witten,glue}. In an underlying string picture the glueball
masses are expected to be proportional to the square root of the string 
tension, like in lattice gauge theory even away from the continuum limit, but 
this is not true here where the 
glueball masses are proportional to the temperature $T$ without any 
$\sqrt{\lambda_T}$ factor. Therefore, in the strong coupling
limit the glueball spectrum does
not appear to be consistent with an underlying string picture, where
the glueballs would come from closed strings, and hence should
have masses proportional to the square root of the string tension \cite{jeff}.
Although this situation could certainly be improved as one goes from 
$g^2N\rightarrow\infty$ to $g^2N\approx 0$, it is strange that the AdS/CFT
approach does not involve a string picture behind the glueballs even in the 
strong coupling limit.

This problem indicates that the definition of the bare coupling should
be reconsidered. In the following we therefore propose the 
(``renormalization'') principle that the scale of the glueball 
masses \cite{Witten,glue},
\begin{equation}
{\rm glueball~masses=const.}~T(1+O(1/\lambda_T)),
\end{equation}
should be the right string scale, so $T$ should be proportional to the
square root of the string tension. For consistency of the supergravity 
calculation around the saddle point, the coupling $\lambda_T$ should still
go to infinity as $T\rightarrow\infty$.
To invoke this ``renormalization'' condition is clearly not possible in 
the leading order, since it would require $\lambda_T$ to be of order one.
However, when fluctuations of the string are included, the string tension
acquires logarithmic corrections \cite{jeff2}. The reason for this 
somewhat unexpected behavior 
is that two of the transverse bosonic world sheet fields become massive, 
whereas the remaining six transverse fields remain massless. The massive fields
then contribute a logarithmic term to the string tension $\Lambda^2$,
\begin{equation}
{\rm string~tension}\equiv\Lambda^2=\frac{8\pi}{27}\lambda_TT^2-4\pi T^2\ln 
\frac{T^2}{\mu^2}\left(1+O(1/\lambda_T)\right).
\label{string}
\end{equation}
Here $\mu$ is an arbitrary scale introduced to regulate the sum over the 
modes of the world sheet fluctuations through the heat kernel for a
Laplace-type operator $\cal O$
\begin{equation}
{\rm tr}~\ln {\cal O}=(4\pi^2\mu^2/e)^{-s}\frac{1}{\Gamma (s)}
\int_0^\infty \frac{dt}{t^{1-s}}{\rm tr}~e^{-t{\cal O}},
\end{equation}
where $s\rightarrow 0$. The factors multiplying $\mu$ have been selected in 
order to simplify the string tension (\ref{string}). The tr ln is thus
evaluated using analytic regularization, and the scale $\mu$ is somewhat 
similar to the arbitrary scale introduced in dimensional regularization.
The two scales $T$ and $\mu$ are a priori of different origin, since $T$ is
the scale at which supersymmetry is broken by the
boundary conditions, whereas $\mu$ is a scale needed to treat the logarithmic
behavior of the massive string modes. 

The equation for the string tension (\ref{string}) can be considered as 
expressing $\Lambda$ in terms of $T$ and $\lambda_T$, but if we have
additional information on the string tension, one can equally well
consider the equation as giving $\lambda_T$ in terms of $T$ and $\Lambda$. Now,
in accordance with the ``renormalization'' principle stated above,
we impose as a boundary condition that the square root of the string 
tension is proportional to the scale $T$ of the glueball masses. Therefore
to leading order in the inverse coupling we require
\begin{equation}
\sqrt{{\rm string~tension}}=\Lambda=c~ T,
\end{equation}
where $c$ is some number which in principle can be fixed if one knows
enough about the glueball masses for higher spins by fitting
the Regge trajectory to these high spins. Hence we get
\begin{equation}
\lambda_T=\frac{27c^2}{8\pi}+27\ln\frac{T}{\mu}\rightarrow 27\ln\frac{T}{\mu}
\rightarrow\infty~~{\rm for}~~T\rightarrow\infty.
\label{asymptotic}
\end{equation}
Therefore the coupling does run with the scale, and the bare coupling goes
logarithmically to infinity (for $T\gg\mu$) when $T\rightarrow\infty$. 
The scale dependent behavior (\ref{asymptotic}) is thus needed in order 
that the glueball masses are proportional to the square root of the string 
tension. This is the main result of this note.

One could ask what should happen
in the end, when all calculations are done some time in the future
(taking into account that
the supergravity approximation must break down for small $\lambda$,
so corrections to the metric should be included) so that it makes sense
to consider also the small coupling.
We would then expect that $\Lambda$, being the square root of the
string tension, becomes the QCD scale, and hence
\begin{equation}
\frac{\Lambda^2}{T^2}\sim e^{-2b/\lambda_{YM}(T)},
\end{equation}
where $\lambda_{YM}(T)\sim b/\ln (T/\Lambda)$ is the QCD coupling at scale 
$T$. Thus, the right physics is obtained by performing the limit
$T\rightarrow\infty$. Then eq. (\ref{string}) would be replaced by
\begin{equation}
\frac{8\pi}{27}\lambda_T-4\pi \ln \frac{T^2}{\mu^2}+{\cal E} \left(
\frac{1}{\lambda_T},\ln \frac{T}{\mu}\right)=\frac{\Lambda^2}{T^2}
\rightarrow 0~{\rm for}~T\rightarrow\infty
\label{future}
\end{equation}
with $\Lambda$ fixed. Here ${\cal E}$ represents the additional $1/\lambda_T$ 
corrections to the string tension divided by $T^2$, coming
from higher order expansions of the string action (including the fermions
as well corrections to the supergravity approximation) used to compute
the fluctuations in the Wilson loop around the saddle point. These 
corrections depend on powers of the inverse coupling 
and could also in general depend on the cutoff. In the interpretation
of eq. (\ref{future}) it is important that $\Lambda$ is renormalization 
group invariant, which defines $\lambda_T$ as a function of $T$ through
eq. (\ref{future}).

In an expansion of the action in orders of fluctuations of the world sheet,
in general the higher order terms do not contribute to the linear term
(string tension)$\times L$, but rather produce terms of order $1/L$ or
smaller. Thus, in general it is not so easy to get contributions to the 
function $\cal E$ from the higher order terms. However, if for example the 
higher order fluctuations produce new mass terms, these can still contribute 
to the string tension. Such terms should involve masses of order $1/\lambda_T$ 
or lower. Also, corrections to the metric may contribute to the string
tension.

Using (\ref{future}) we then get
\begin{equation}
\lambda_T\approx 27\ln\frac{T}{\mu}-\frac{27}{8\pi}{\cal E} \left(
\frac{1}{\lambda_T},\ln \frac{T}{\mu}\right)~~{\rm for}~~T\rightarrow \infty.
\label{result}
\end{equation}
This is an implicit equation for $\lambda_T$. Computing more and more terms
in $\cal E$ change the functional dependence of $\lambda_T$ on $\ln (T/\mu)$.
Then, if everything goes well, eq. (\ref{result}) should 
have a solution \cite{gr}
\begin{equation}
\lambda_T\approx b/\ln (T/\mu)~~{\rm for}~~T\rightarrow \infty,
\label{as}
\end{equation}
approaching zero in this limit. 

It should be emphasized that a solution of the type (\ref{as}) cannot in
general be the only solution of (\ref{future}). For example, if $\cal E$ 
only contains a finite number of significant terms and does not depend 
on $\ln (T/\mu)$, then e.g. a strong, logarithmically divergent
coupling is always a solution, because for this particular case $\cal E$
can be ignored for the strong coupling in eq. (\ref{future}), since 
for $\lambda_T\rightarrow\infty$ the $1/\lambda_T$ dependence of $\cal E$ is
insignificant relative to the leading terms on the left hand side of
eq. (\ref{future}). This is valid more generally if $\cal E$ is analytic in 
$1/\lambda_T$, which presumably is equivalent to having no phase 
transition \cite{gr} in going from strong to weak coupling. In this
case eq. (\ref{string}) is a rudimentary version of eq. (\ref{future}).
Also, if $\cal E$ only has a finite number $q$ of significant 
terms, eq. (\ref{future}) is a polynomial equation of order $q$, and hence 
can have $q$ solutions, some of which may be invalid because they are complex.
If there are an infinite number of terms in $\cal E$ the situation is 
of course quite different. A phase transition may occur so that $\cal E$ is
not analytic in $1/\lambda_T$. 

Actually $\cal E$ does not have to be terribly complicated to produce an
answer which looks much like the right one, as the following $hypothetical$
example shows. Suppose we obtain
\begin{equation}
{\cal E}=\frac{k}{\lambda_T}
\label{example}
\end{equation}
in a calculation where corrections to the metric are included, so that
it makes sense to consider small values of $\lambda_T$.
Here $k$ is a positive constant, and it is assumed
further that higher order terms are absent or have very small coefficients and 
can be ignored. Then from (\ref{future}) we find that $\lambda_T$ satisfies
\begin{equation}
\lambda_T^2-27\ln (T/\mu)\lambda_T+27k/8\pi=O(\Lambda^2/T^2)\approx 0,
\end{equation}
where $\Lambda$ is fixed, so that the right hand side of this equation
is sub-logarithmic, of order $1/T^2$. Thus we have the two solutions
\begin{equation}
\lambda_T\approx 27\ln (T/\mu)+O(1/\ln (T/\mu))~~{\rm and}~~k/
[8\pi\ln (T/\mu)]+O(1/\ln^3 (T/\mu)),
\label{last}
\end{equation}
corresponding to a strong coupling\footnote{This is not the same as the
bare strong coupling in eq. (\ref{asymptotic}), which was derived in the 
strong coupling regime where $\Lambda=cT$, in contrast to the
fixed $\Lambda$ behavior relevant when all corrections
are included. However, the logarithmic divergence of the strong coupling
is exactly the same in the two cases.} and to the right logarithmically
decreasing behavior of the asymptotically
free QCD coupling, respectively, provided we identify the so far arbitrary 
scale $\mu$ with the QCD scale $\Lambda$. If $k=8\pi b$ we would then get the 
right answer for the weak coupling. Since there are two solutions
for the coupling, we have the option of taking the weak coupling $\lambda_T$
as the right solution. Of course, this assumes that the formula (\ref{example})
really includes corrections such that the metric makes sense even for
small $\lambda_T$.

This is certainly a fictitious example, and 
higher order terms in $1/\lambda_T$ could play an important 
role. However, the example shows that the strong$\rightarrow$
weak coupling transition could happen in a relatively simple way, and  
it would anyhow be of interest to compute the next $1/\lambda_T$ order, 
since it could give the right functional dependence of the coupling on the 
logarithm. It would then be interesting to see how far the coefficient of 
the inverse 
logarithm is from the right value $b$. We do not expect to get the right
value of $b$, of course, before the problems connected with the break down of 
the supergravity approximation at small $\lambda_T$ have been settled. This 
would presumably imply that the coefficient $k$
does not have the right value $8\pi b$, or that there is no $1/\lambda_T$
correction ($k=0$).
Further, it could be that $k$ is not really a constant, but depends on 
the cutoff $T$. For example, if $k=K\ln (T/\mu)$, 
where $K$ is a positive constant, corresponding to ${\cal E}=K\ln (T/\mu)/
\lambda_T$, we would again get the strong coupling limit
as in (\ref{last}) but the other solution would be a constant $K/8\pi$. 
\vskip0.3cm
In conclusion, the main points of this paper are:
\begin{itemize}
\item Proportionality between the glueball masses and the square root of the
string tension requires the bare strong coupling to run with $T$. If, in
contrast, the coupling is considered as an arbitrary parameter, there is no
underlying string picture of the glueballs in the strong coupling limit. 	
\item In the end, when all corrections are included, we get an equation for 
the coupling which has a number of solutions (if there is no phase transition
between weak and strong coupling). One of these solutions is hopefully the 
right one exhibited in eq. (\ref{scaling}), but another one is presumably the 
running bare strong coupling (\ref{asymptotic}). If there is a phase 
transition, these two solutions would be on ``different branches''. 
\item There exists a very simple case which exhibits the strong and
weak coupling, namely the behavior (\ref{example}). 
\item There exists a curious relation between the bare strong coupling and
the asymptotically free one, namely
\begin{equation}
\lambda_T\approx 27b/\lambda_{YM}+27c^2/8\pi\sim
27b/\lambda_{YM}~~{\rm for}~~T\rightarrow\infty,
\end{equation}
provided $\mu$ is identified with $\Lambda_{QCD}$.
This might indicate that the regulator theory (with $\lambda_T
\rightarrow\infty$)
has a dual relation to QCD, so that the former is a strong coupling version
of the latter, and vica versa. Of course, it could also be that this relation
is purely accidental.
\end{itemize}

We end by the remark that the running coupling $\lambda_T\propto \ln (T/\mu)$
discussed in this note can be understood as a renormalization needed
in strings with extrinsic curvature \cite{extr}. It was pointed
out in ref. \cite{jeff2} that when world sheet fluctuations are included,
the string in the AdS/CFT approach becomes rigid, and hence it is well known 
that there are renormalizations \cite{extr}. The physical situation
in the present case is, however, different from the one discussed in 
the literature on strings with extrinsic curvature.

\end{document}